\documentclass[12pt,reqno]{article}
\usepackage[lmargin=2cm, rmargin=2cm]{geometry}
\usepackage{amsfonts}
\usepackage{amssymb}
\usepackage{enumerate}
\usepackage{amsthm}
\usepackage{array}
\usepackage{epsfig}
\usepackage[small,bf]{caption}
\usepackage{appendix}
\usepackage{dsfont}
\usepackage{hyperref}
\usepackage{amsmath}
\usepackage{graphicx}
\usepackage{epstopdf}
\usepackage{blkarray}
\usepackage{cite}

\allowdisplaybreaks \numberwithin{equation}{section}

\begin{document}

\begin{titlepage}
 \thispagestyle{empty}

\begin{flushright}
 \end{flushright}

 \begin{center}

 \vspace{30mm}

     { \LARGE{\bf  {De Finetti theorems and entanglement in large-N theories and gravity}}}

     \vspace{40pt}

\Large{{\bf Javier M. Mag\'an}} \\[8mm]
{\small\slshape Instituto Balseiro, Centro At\'omico Bariloche \\
S.~C. de Bariloche, R\'io Negro, R8402AGP, Argentina\\

\vspace{5mm}

{\upshape\ttfamily javier.magan@cab.cnea.gov.ar}\\[3mm]}

\vspace{8mm}

     \vspace{10pt}

    \vspace{10pt}

\date{\today}

\end{center}

\begin{abstract}
The de Finetti theorem and its extensions concern the structure of multipartite probability distributions with certain symmetry properties, the paradigmatic original example being permutation symmetry. These theorems assert that such symmetric distributions are well approximated by convex combinations of uncorrelated ones. In this article, we apply de Finetti theorems to quantum gravity theories, such as the Sachdev-Ye-Kitaev (SYK) model or large-N vector and gauge theories. For SYK we put recent studies of information/entanglement dynamics in a general and rigorous basis. For vector and gauge theories, we find a gauge invariant operator whose expectation value provides the leading term in the entanglement entropy in all states close enough to a given classical state. These results can be unified through a generic statement about the nature of Schmidt decompositions and decoherence in large-N theories. In the reverse direction, we extend de Finetti theorems in various ways and provide an independent approach to the theorems only based on the large-N properties of the gauge invariant coherence group.
\end{abstract}

 \vspace{10pt}
\noindent

\end{titlepage}

\thispagestyle{plain}

\tableofcontents

\baselineskip 6 mm

\newpage

\section{Introduction}\label{secI}

In this article we discuss aspects of the state/entanglement structure in quantum gravity. This has been an active area of research ever since Ref.\cite{bombelli} appeared, where spatial entanglement was argued to be a source of black hole entropy. It has now a prominent role since the discovery in Ref.\cite{taka} that natural geometrical variables, such as minimal area surfaces, provide the amount of boundary spatial entanglement present in the context of AdS/CFT \cite{adscft}. Many further aspects and implications have been developed since then, see the review \cite{takareview}.

Most of these discussions consider spatial entanglement entropy \footnote{Some exceptions are \cite{usfreeblack,bala1,bala2,sahakian}.}, obtained when tracing out part of space. But in AdS/CFT, the field theory is a large-N gauge theory, and we would like to understand the state/entanglement structure in color space, as well as implications of the large-N limit in such structure. There are several motivations. First, there are holographic dualities with no space in the non-gravitational description, namely matrix quantum mechanics \cite{matrix}, and AdS2/SYK \cite{kitaev,remarks}. In these cases, the connection between entanglement and geometry has to be related to entanglement within `internal' degrees of freedom. Second, the large-N limit \cite{largeN} is a fundamental input in AdS/CFT to derive approximate locality in the gravitational theory, see \cite{kabat,kiriakos}. Caracteristic aspects of entanglement in this limit are expected as well, probably connected with bulk locality \cite{kabat2}. Another motivation comes from understanding the origin of the Ryu-Takayanagi formula \cite{taka,aitor}, a peculiarity of holographic theories and error correcting codes \cite{daniel}. Also, the dynamics in color space is fundamental for black hole physics \cite{matrixpol,susskind,us1,bound,usfreeblack}. Finally, from a general standpoint, most quantum gravity theories have some built-in large-N structure, controlling the classical gravity limit. Understanding some generic entanglement behavior in this classical limit is an important question.

In this article we approach such problems by joining and expanding results and ideas from \cite{definc,definq,facile,yaffe,usfreeblack,usperm}. The approach rests upon symmetries, permutation symmetry for SYK, and gauge symmetry for gauge theories, allowing the use of de Finetti theorems \cite{definc,definq,facile}. An independent approach is developed based on the structure of coherent states in large-N theories \cite{yaffe}.

The results are generic formulas for reduced density matrices and entanglement entropies, which can be traced to a statement about Schmidt decompositions in large-N theories. For SYK we rigorously prove the entanglement dynamics inferred in Ref. \cite{usfreeblack}, where a de Finetti like theorem was derived by direct computation. In the context of the Ryu-Takanayagi formula \cite{taka,aitor}, we show how to construct observables whose expectation values provide the leading term of entanglement entropies in subspaces close to a given classical state.

On the way we expand on the subject of de Finetti theorems,  generalizing the theorem in various directions and providing the measures appearing in the theorem in several cases. We also show that large-N factorization suggests that in quantum gravity symmetry constrains are more restrictive than expected for other cases.

\section{Classical and quantum de Finetti theorems}\label{secII}

The classic theorem \cite{definc} considers probability distributions $p_{N}(x_{1},x_{2},\cdots ,x_{N})$ invariant under permutations of its variables. Such distributions are said to be M-exchangeable if there is another permutation symmetric distribution $p_{M}(x_{1},x_{2},\cdots ,x_{M})$ satisfying $p_{N}(x_{1},x_{2},\cdots ,x_{N})=\textrm{tr}_{M-N}\,p_{M}(x_{1},x_{2},\cdots ,x_{M})$. A distribution is infinitely exchangeable if there exists such extended distribution for any $M$. The theorem asserts that infinitely exchangeable distributions are convex combinations of identically distributed uncorrelated ones:
\begin{equation}\label{cdefin}
p_{N}(x_{1},x_{2},\cdots ,x_{N})=\int d\mu [p (x)]\,p(x_{1})p(x_{2})\cdots p(x_{N})\;,
\end{equation}
where $ \mu [p(x)]$ is a measure on the space of probability distributions of one variable (see \cite{prueba} for a simple proof). If $p_{N}$ is only $M$-exchangeable, the convex combination is an approximation with an error depending on the ratio $N/M$. Theorem~(\ref{cdefin}) is a paradigmatic result, showing how global symmetry can constrain the state structure. 

Quantum extensions have received attention recently, see \cite{definq,facile} and references therein. For infinitely exchangeable permutation symmetric states $\rho^{k}$ in the tensor product of $k$ systems:
\begin{equation}\label{qdefins}
\rho^{k}=\int \sigma^{\otimes k}\,dm [\sigma]\;,
\end{equation}
where $m [\sigma]$ is a measure on the space of density matrices of one system. A simple proof goes by noticing that in permutation invariant states, complete sets of measurements define permutation invariant probability distributions, to whom the classical theorem~(\ref{cdefin}) applies.

More interestingly, a generalized de Finetti theorem was proven in Ref \cite{facile}. It starts with an unimodular group $G$, i.e a group with an invariant Haar measure $\mu (g)$ and square integrable representations $\mathcal{A}$ :
\begin{equation}
\int\limits_{G}\vert\langle\psi\vert \mathcal{A}(g)\vert\varphi\rangle\vert^{2}\,d\mu (g)<\infty \,\,\,\,\,\,\,\,\,\forall \vert\varphi\rangle,\vert\psi\rangle \in\mathcal{A}\;,
\end{equation}
satisifying a version of Schur's lemma:
\begin{equation}
\int\limits_{G}\langle\beta\vert \mathcal{A}(g)^{\dagger}\vert\alpha\rangle\langle\gamma\vert \mathcal{A}(g)\vert\delta\rangle \,d\mu (g)=\frac{\langle\gamma\vert\alpha\rangle\langle\beta\vert\delta\rangle}{d_{\mathcal{A}}}\;,
\end{equation}
where $d_{\mathcal{A}}$ is called the formal degree of the representation. The previous relation implies the usual irreducibility condition:
\begin{equation}
d_{\mathcal{A}}\int\limits_{G}\mathcal{A}(g)\vert\psi\rangle\langle\psi\vert \mathcal{A}(g)^{\dagger} \,d\mu (g)=\mathds{1}_{\mathcal{A}}\,\,\,\,\,\,\,\,\,\forall \,\vert\psi\rangle \in\mathcal{A}\;.
\end{equation}
In such $G$, consider three irreducible representations $\mathcal{A}$,  $\mathcal{B}$ and $\mathcal{C}$, with $\mathcal{C}\subset \mathcal{A}\otimes\mathcal{B}$. The theorem asserts that if $\mathcal{X}$ is a subspace of $\mathcal{A}$, and defining $\vert\psi_{g}^{\mathcal{X}} \rangle\equiv g\vert\psi^{\mathcal{X}}\rangle$,  with $\vert\psi^{\mathcal{X}}\rangle\in\mathcal{X}$, then for any $\vert\Psi\rangle \in \mathcal{C}$ there is a measure $m (g)$ on $G$ satisfying:
\begin{equation}\label{qdefing}
\Vert \textrm{Tr}_{\mathcal{B}}\vert\Psi\rangle\langle\Psi\vert -\int \vert\psi_{g}^{\mathcal{X}}\rangle\langle\psi_{g}^{\mathcal{X}}\vert \,dm (g)\Vert \leq \epsilon (\mathcal{X})\equiv 2 \sqrt{1-\delta (\mathcal{X})} \;,
\end{equation}
where $\Vert A \Vert \equiv\frac{1}{2}\textrm{Tr}\sqrt{A^{\dagger}A}$ and:
\begin{equation}\label{delta}
\delta (\mathcal{X})\equiv \underset{\vert\phi\rangle\in\mathcal{B}}{\textrm{sup}} \,\frac{d_{\mathcal{B}}}{d_{\mathcal{C}}}\,\textrm{Tr}\,[P_{\mathcal{C}}\,(P_{\mathcal{X}}\otimes\vert\phi\rangle\langle\phi\vert)\,]\;.
\end{equation}
The proof just uses the triangle inequality for trace distances and Schur's lemma. Equation~(\ref{delta}) implies
\begin{equation}
\mathcal{X}\subset\mathcal{X'}\Rightarrow\delta (\mathcal{X})\leq \delta (\mathcal{X'})\;,
\end{equation}
saying that taking smaller subspaces worsens the approximation. Given the irreducibility of representations it also implies
\begin{equation}
\mathcal{X}=\mathcal{A}\Rightarrow \delta (\mathcal{X})=1\;.
\end{equation}
We conclude that $\delta (\mathcal{X})\leq 1$, so that $\epsilon (\mathcal{X})\geq 0$, with equality when $\mathcal{X}=\mathcal{A}$. In simple words, the theorem asserts that given a subspace $\mathcal{X}$, there exists a convex combination of vectors $\vert\psi_{g}^{\mathcal{X}} \rangle\langle\psi_{g}^{\mathcal{X}}\vert\equiv g\vert\psi^{\mathcal{X}}\rangle\langle\psi^{\mathcal{X}}\vert g$ that faithfully approximates the reduced state.  The approximation error decreases as we increase the subspace $\mathcal{X}$, or as we increase the size of $\mathcal{B}$ relative to $\mathcal{C}$. Notice also that if the $\mathcal{X}$ subspace consists only on one vector $\vert \varphi\rangle \in\mathcal{A}$, and if there is a vector $\vert\psi\rangle\in \mathcal{B}$ such that $\vert 0\rangle\otimes\vert\psi\rangle \in \mathcal{C}$, then $\delta (\mathcal{X})=d_{\mathcal{B}}/d_{\mathcal{C}}$. In such case it was proven in \cite{qdefinimprovebound} that the error decreases to $\epsilon (\mathcal{X})= 2 (1-\delta (\mathcal{X}))$.

Theorem~(\ref{qdefing}) again shows how global symmetry conditions, like belonging to some irreducible representation of the group, constrains reduced states. As an application, one can obtain~(\ref{qdefins}) from~(\ref{qdefing}) by choosing symmetric irreducible representations of unitary groups, see \cite{facile}.

The theorem can be extended in three simple ways, as needed for gauge theories. First, notice that the theorem should only care about the global representation $\mathcal{C}$ and the local $\mathcal{A}$. In such case, if $\mathcal{X}$ is a subspace of $\mathcal{A}$, and defining $\vert\psi_{g}^{\mathcal{X}} \rangle\equiv g\vert\psi^{\mathcal{X}}\rangle$,  with $\vert\psi^{\mathcal{X}}\rangle\in\mathcal{X}$, then for any $\vert\Psi\rangle \in \mathcal{C}$ there is a measure $m (g)$ on $G$ satisfying:
\begin{eqnarray}\label{qdefingg}
&\delta (\mathcal{X})\equiv \underset{\mathcal{B}\,/\, \mathcal{A}\otimes\mathcal{B}=\mathcal{C}\oplus \cdots}{\textrm{sup}}\,\,\lbrace\underset{\vert\phi\rangle\in\mathcal{B}}{\textrm{sup}} \,\frac{d_{\mathcal{B}}}{d_{\mathcal{C}}}\,\textrm{Tr}\,[P_{\mathcal{C}}\,(P_{\mathcal{X}}\otimes\vert\phi\rangle\langle\phi\vert)\,]\rbrace &\nonumber \\
&\Vert \textrm{Tr}_{\mathcal{B}}\vert\Psi\rangle\langle\Psi\vert -\int \vert\psi_{g}^{\mathcal{X}}\rangle\langle\psi_{g}^{\mathcal{X}}\vert \,dm (g)\Vert \leq \epsilon (\mathcal{X})\equiv 2 \sqrt{1-\delta (\mathcal{X})}&  
\;.
\end{eqnarray}
In simple words, $\mathcal{C}$ might appear in $\mathcal{A}\otimes\mathcal{B}$ for different $\mathcal{B}$, and we should choose the $\mathcal{B}$ with smallest error. This minimization over $\mathcal{B}$ improves the error bound in~(\ref{qdefing}). Generically, we expect the smallest error for the $\mathcal{B}$ with biggest dimension.

Another generalization appears when considering a reducible representation $\mathcal{C}$. Writing it as a sum $\mathcal{C}=\oplus \mathcal{C}_{\alpha}$ of irreducible representantions $\mathcal{C}_{\alpha}\subset \mathcal{A}\otimes\mathcal{B}$, a state $\vert\Psi\rangle\in\mathcal{C}$ reads $\vert\Psi\rangle=\sum\limits_{\alpha}\psi_{\alpha}\vert\psi_{\alpha}\rangle$, with $\vert\psi_{\alpha}\rangle\in\mathcal{C}_{\alpha}$. In case the observables are linear combinations of group elements, each $\alpha$ behaves as a superselection sector:
\begin{equation}
\rho_{\mathcal{C}}=\sum\limits_{\alpha}\vert\psi_{\alpha}\vert^{2}\vert\Psi_{\alpha}\rangle\langle\Psi_{\alpha}\vert = \sum\limits_{\alpha}p_{\alpha}\vert\Psi_{\alpha}\rangle\langle\Psi_{\alpha}\vert\;,
\end{equation}
Using $\Vert A+B \Vert\leq \Vert A\Vert+\Vert B \Vert$, we conclude there exist measures $m_{\alpha} (g)$ on $G$, subspaces $\mathcal{X}_{\alpha}$ and states $\vert \psi^{\alpha}_{g}\rangle=\mathcal{A}(g)\vert \psi^{\alpha}\rangle$, with $\vert \psi^{\alpha}\rangle\in \mathcal{X}_{\alpha}$ satisfying:
\begin{equation}\label{qdefint1}
\Vert \textrm{Tr}_{\mathcal{B}}\vert\Psi\rangle\langle\Psi\vert -\sum\limits_{\alpha}p_{\alpha}\int \vert\psi^{\alpha}_{g}\rangle\langle\psi^{\alpha}_{g}\vert dm_{\alpha} (g)\Vert \leqslant \sum\limits_{\alpha}2p_{\alpha} \sqrt{1-\delta (\mathcal{X}_{\alpha})}\;.
\end{equation}
This states that, when combining superselection sectors, the error is the average over the errors in each sector.

Finally we can consider mixed states $\rho_{\alpha}$ in each sector. Diagonalizing $\rho_{\alpha}=\sum\limits_{i_{\alpha}}\lambda_{i_{\alpha}}\vert i_{\alpha}\rangle\langle i_{\alpha}\vert$ we can apply~(\ref{qdefing}) to each $\vert i_{\alpha}\rangle$ and conclude that the error bound does not change when considering mixed states.

\section{Entanglement dynamics in SYK}\label{secIII}

SYK models \cite{kitaev,remarks} contain $N$ Majorana fermions interacting through random k-body interactions:
\begin{equation}\label{SYK2}
H= i^{k}\sum\limits_{1\leq i_{1}<\cdots<i_{2k}\leq N}J_{i_{1}\cdots i_{2k}}\chi_{i_{1}}\cdots \chi_{i_{2k}}\;.
\end{equation}
Each term contains $2k$ Majorana operators and the couplings are real random numbers with zero mean and variance equal to $J$. A different version is
\begin{equation}\label{SYK}
H=\sum\limits_{1\leq i_{1}<\cdots<i_{k}\leq N \atop 1\leq j_{1}<\cdots<j_{k}\leq N}J_{i_{1}\cdots i_{k};j_{1}\cdots j_{k}}c^{\dagger}_{i_{1}}\cdots c^{\dagger}_{i_{k}}c_{j_{1}}\cdots c_{j_{k}}\;,
\end{equation} 
where $J_{i_{1}\cdots i_{k};j_{1}\cdots j_{k}}$ are real random numbers and hermiticity requires $J_{i_{1}\cdots i_{k};j_{1}\cdots j_{k}}=J_{j_{1}\cdots j_{k};i_{1}\cdots i_{k}}$. Each term of the Hamiltonian contains $k$ annihilation and $k$ creation operators. For an even number of Majoranas, the first Hamiltonian can be written in terms of $N/2$ Dirac fermions. The Hilbert space of both theories can then be expanded by eigenvectors of products of number operators $\vert a_{1}a_{2}\cdots a_{N}\rangle$, with $a_{1}a_{2}\cdots a_{N}={0,1}$.

There are various motivations to study these models. They are excellent models for discussions of quantum chaos and thermalization \cite{kbody,kitaev,usfree,usperm}, since dissipative phenomena can be treated analytically. They were shown to have holographic duals \cite{kitaev}, and saturate the chaos bound \cite{bound}, see \cite{remarks} for a complete discussion. Besides, they could potentially be created in the lab \cite{labSYK}. Parallely, as developed in Ref. \cite{usfreeblack,usperm}, they are good toy models where one might extract generic conclusions for other large-N theories.

Consider an initial state with the first $m$ fermions excited:
\begin{equation}
\vert\psi_{\textrm{in}}\rangle =\vert 1_{1}\cdots 1_{m}0_{m+1}\cdots 0_{N}\rangle\;.
\end{equation}
Unitarily evolving with~(\ref{SYK2}) or~(\ref{SYK}) and averaging:
\begin{equation}
\overline{\rho(t)}=\overline{U(t)\vert\psi_{\textrm{in}}\rangle\langle\psi_{\textrm{in}}\vert U^{\dagger}(t)}=\sum_{a_{1}a_{2}\cdots a_{N}}p_{a_{1}a_{2}\cdots a_{N}}(t)\vert a_{1}a_{2}\cdots a_{N}\rangle\langle a_{1}a_{2}\cdots a_{N}\vert\;,
\end{equation}
one obtains a state specified by the average probabilities $p_{a_{1}a_{2}\cdots a_{N}}(t)\equiv \overline{\vert\langle a_{1}a_{2}\cdots a_{N}\vert U(t)\vert \psi_{\textrm{in}}\rangle\vert^{2}}$ of Fock basis states. This was proven in \cite{usperm}, by showing the following relation for the Hamiltonian statistics in the Fock basis:
\begin{equation}\label{averageprod}
\overline{(H^{n})_{ji}(H^{m})_{ij'}}= \sum\limits_{k_{1}k_{2}\cdots k_{n+m-2}}\overline{H_{jk_{1}}H_{k_{1}k_{2}}\cdots H_{k_{n-1}i}H_{ik_{n}}\cdots H_{k_{n+m-2}j'}}\propto \delta_{jj'}\;,
\end{equation}
where $ (H^{n})_{ji}\equiv\langle j\vert H^{n}\vert i\rangle$.

By the same reasoning, any reduced state is also diagonal on average. In particular, the reduced density matrix of the initially excited fermions reads:
\begin{equation}
\overline{\rho^{A} (t)}=\sum_{a_{1}a_{2}\cdots a_{m}}p_{a_{1}a_{2}\cdots a_{m}}(t)\vert a_{1}a_{2}\cdots a_{m}\rangle\langle a_{1}a_{2}\cdots a_{m}\vert\;,
\end{equation}
a relation which shows the strength of decoherence in SYK. Since $\rho^{A}_{\textrm{in}}=\vert 1_{1}\cdots 1_{m}\rangle\langle 1_{1}\cdots 1_{m}\vert$ and the unitary evolution display permutation symmetry, $p_{a_{1}a_{2}\cdots a_{m}}(t)$ is permutation invariant, so the classical theorem~(\ref{cdefin}) applies to it:
\begin{equation}
\overline{\rho^{A}}\simeq\sum_{a_{1}a_{2}\cdots a_{m}}\int d\mu [p,t]\,p(a_{1})p(a_{2})\cdots p(a_{m})\vert a_{1}a_{2}\cdots a_{m}\rangle\langle a_{1}a_{2}\cdots a_{m}\vert\;,
\end{equation}
where $\mu [p,t]$ is a time dependent measure on the space of binary distributions \footnote{Notice that a binary distribution is just defined by a real number $p\in [0,1]$.}. Defining $\rho (p)=p\vert 0\rangle\langle 0\vert +(1-p)\vert 1\rangle\langle 1\vert$:
\begin{equation}\label{sykd}
\overline{\rho^{A}(t)}\simeq\int d\mu [p,t]\rho (p)^{\otimes m}\;.
\end{equation}
showing that reduced states in SYK are close to separable states at all times. There is little quantum entanglement in the large-N limit, the connected correlators being controlled by the classical measure $\mu [p, t]$. Notice that for one degree of freedom
\begin{equation}\label{ro1}
\overline{\rho^{1}(t)}\simeq\int d\mu [p,t]\rho (p)\;,
\end{equation}
so $\mu [p,t]$, controlling higher point correlation functions, appears at the local level, implying very non-trivial relations between high and low point correlation functions. This furnishes a deeper version of the global/local relation inferred in \cite{usfreeblack}.

Finally, entropy evolution satisfies\footnote{ Relation~(\ref{sykd2}) is to be understood after a discrezation of the measure $\mu [p,t]$ is taken, so that Shannon entropy is well defined. Such discretization is physically well motivated, in the sense of indistinguishability of quantum states.}
\begin{equation}\label{sykd2}
m\int d\mu [p,t]S(\rho (p))\leqslant S(\rho^{A}(t))\leqslant m\int d\mu [p,t]S(\rho (p))+S(\mu [p,t])\;,
\end{equation}
where we remark that $S(\mu [p,t])$ does not increase with $m$. Since for sufficiently small subsystems $A$, unitary evolution drives the reduced state towards the maximally mixed state $\rho_{\textrm{mix}}\equiv\frac{1}{\Omega_{A}}\sum\limits_{i_{A}}\vert i_{A}\rangle\langle i_{A}\vert$, the previous relation can be used to bound the quantum distance to thermality:
\begin{equation}\label{ent1}
S(\rho^{A},\rho_{\textrm{mix}})\equiv \textrm{Tr} \rho^{A}(\log \rho^{A}-\log \rho_{\textrm{mix}})=\log \Omega_{A} -S(\rho^{A})\geq \log \Omega_{A}-m\int d\mu [p,t]S(\rho (p))\;,
\end{equation}
where we have used the relative entropy $S(\rho^{A},\rho_{\textrm{mix}})$ as a notion of quantum distance \cite{petz}. Eqaution~(\ref{ent1}) shows that large and small subsystems share the same characteristic time scales, those of $\mu [p,t]$. 

Finally, taking into account large-N factorization \cite{largeN}, a generic feature of large-N theories asserting that n-point correlation functions $\mathcal{O}_{1}\mathcal{O}_{2}\cdots \mathcal{O}_{n}$ factorize in the large-N limit, it follows:
\begin{equation}
\lim_{N\rightarrow\infty}\rho^{A}(t)=(\rho^{1}(t))^{\otimes m}\;,
\end{equation}
implying:
\begin{equation}
\lim_{N\rightarrow\infty}S(\mu [p,t])=0 ~~~~~~~~~~~~~~S(\rho^{A}(t))=mS(\rho^{1}(t))\;,
\end{equation}
verifying the claims in \cite{usfreeblack} and suggesting that for gravity applications, symmetry constraints force the measure $\mu [p]$ to have vanishing entropy in the thermodynamic limit.

\section{State structure in large-N theories}\label{secIV}

To discuss vector and gauge theories, we follow Ref. \cite{yaffe}, whose objective was to show why theories with a large-N limit are `classical'. The challenge was to find a classical phase space within the Hilbert space, and show that quantum dynamics sources classical dynamics in such phase space. This was achieved by finding generalized coherent states. Let us briefly review the logic here.

In large-N theories, the Hilbert space $\mathcal{H}_{N}$ depends on $N$. Ref. \cite{yaffe}  shows that for such theories there exists a unimodular group $G$ (as defined in~(\ref{secII})), named the `coherence group', irreducible representations $\mathcal{R}_{N}^{G}$ acting on each $\mathcal{H}_{N}$, and a base state $\vert 0_{N}\rangle$ satisfying,
\begin{equation}
d_{N}\int\limits_{G}\mathcal{R}_{N}^{G}(g)\vert 0_{N}\rangle\langle 0_{N}\vert \mathcal{R}_{N}^{G}(g)^{\dagger} d\mu (g)\equiv d_{N}\int\limits_{G}\vert g_{N}\rangle\langle g_{N}\vert d\mu (g)=\mathds{1}_{\mathcal{H}_{N}}\;,
\end{equation}
where $d_{N}$ is the formal degree of $H_{N}$ and $\mu (g)$ the Haar measure on $G$. $\vert g_{N}\rangle\in \mathcal{H}_{N}$ is an overcomplete (gauge invariant) basis of the Hilbert space, the coherent states. We remark the essential point that the coherence group does not depend on $N$, just the associated irreducible representation $\mathcal{R}_{N}^{G}$.

As happens when taking the limit $\hbar\rightarrow 0$ in non-relativistic quantum mechanics, not every operator has a sensible $N\rightarrow \infty$ limit. It is natural to define the class of `classical operators' $\textbf{A}$, such that $\langle g_{N}\vert A\vert g'_{N}\rangle/\langle g_{N}\vert g'_{N}\rangle$ has a finite limit. Since $\textbf{A}$ is the not the full operator algebra, it will sometimes fail to distinguish different coherent states, defining an equivalence class on $G$. Two coherent states are said to be equivalent $g\sim g'$ if
\begin{equation}
\lim_{N\rightarrow \infty}\langle g_{N}\vert A\vert g_{N}\rangle=\lim_{N\rightarrow \infty}\langle g'_{N}\vert A\vert g'_{N}\rangle\;.
\end{equation}
Ref. \cite{yaffe} shows that inequivalent coherent states become orthogonal in the $N\rightarrow \infty$ limit, implying that classical operators cannot connect inequivalent states. Lastly, it is shown that the Hamiltonian belongs to the class of classical operators and how to construct a classical phase space from the space of equivalent classes of coherent states.

Let us derive from these properties a de Finetti like approximation. Consider a large-N theory and a state whose modular Hamiltonian $H$ is a classical operator (thermal density matrices are examples). Expanding the operator in the coherent state basis, and since a classical operator cannot connect inequivalent coherent states:
\begin{eqnarray}\label{global}
H&=&d_{N}^{2}\int\limits_{G} \langle g_{N}\vert  H \vert g_{N}'\rangle     \vert g_{N}\rangle\langle g'_{N}\vert   d\mu (g)d\mu (g')\xrightarrow[N\rightarrow \infty]{}d_{N}^{2}\int\limits_{G} H(g)     \vert g_{N}\rangle\langle g_{N}\vert   d\mu (g)\nonumber \\
e^{-H}&=&d_{N}^{2}\int\limits_{G} \langle g_{N}\vert  e^{-H} \vert g_{N}'\rangle     \vert g_{N}\rangle\langle g'_{N}\vert   d\mu (g)d\mu (g')\xrightarrow[N\rightarrow \infty]{}d_{N}^{2}\int\limits_{G} e^{-H}(g)    \vert g_{N}\rangle\langle g_{N}\vert   d\mu (g)
\;.
\end{eqnarray}
This relation states that global coherent states universally diagonalize states with classical modular Hamiltonians. Now, dividing the system in two (we consider an explicit example below), the subsystems modular Hamiltonians are also classical operators with respect to each subsystem coherent state basis. Intuitively, reducing can only make things less sensitive to off-diagonal physics (jumps between different coherent states). Mathematically, global coherent states can be expressed in the tensor product basis of local coherent states:
\begin{equation}
\vert g_{N}\rangle=\int\limits_{G}d\mu (g)d\mu (g')\psi (g,g')\vert g_{A}\rangle\otimes\vert g_{B}'\rangle 
\end{equation}
If the operator $\mathcal{O}=H_{A}\otimes \mathds{1}_{B}$, where $H_{A}$ is susbsystem's $A$ modular Hamiltonian, is to be a classical operator with respect to global coherent states (remember that $H$ is assumed to be a classical operator), and since $\mathds{1}_{B}$ is classical with respect to $B$'s coherent states, then $H_{A}$ is classical with respect to $A$'s coherent states:
\begin{eqnarray}\label{redco}
H_{A}&=&d_{A}^{2}\int\limits_{G} \langle g_{A}\vert  H_{A} \vert g_{A}'\rangle     \vert g_{A}\rangle\langle g'_{A}\vert   d\mu (g)d\mu (g')\xrightarrow[N\rightarrow \infty]{}d_{A}^{2}\int\limits_{G} H_{A}(g)     \vert g_{A}\rangle\langle g_{A}\vert   d\mu (g)\nonumber \\
e^{-H_{A}}&=&d_{A}^{2}\int\limits_{G} \langle g_{A}\vert  e^{-H_{A}} \vert g_{A}'\rangle     \vert g_{A}\rangle\langle g'_{A}\vert   d\mu (g)d\mu (g')\xrightarrow[N\rightarrow \infty]{}d_{A}^{2}\int\limits_{G} e^{-H_{A}}(g)    \vert g_{A}\rangle\langle g_{A}\vert   d\mu (g)
\;.
\end{eqnarray}
In large-N theories, reduced states are well approximated by convex combinations of the subsystem coherent state basis, a de Finetti approximation where the $\mathcal{X}$ subspace is just the base state $\vert 0_{N}\rangle$. From a physical perspective, the previous relation shows the strength of decoherence in large-N theories, when studied in the coherent state basis.

Let's arrive at the same conclusion by using the theorem. We begin with vector models. The operators in these theories are positions $\hat{x}_{i}$ and momentum $\hat{p}_{i}$, for $i=1,\cdots , N$. The Hamiltonian and the Hilbert space $\mathcal{H}_{N}$ are $O(N)$ invariant. As shown in \cite{yaffe}, there is a coherence group $G$ and irreducible representations $\mathcal{R}_{N}^{G}$ for every $N$ (the Lie algebra of the group are linear combinations of $\sum \hat{x}_{i}\hat{x}_{i}$ and $\sum (\hat{x}_{i}\hat{p}_{i}+\hat{p}_{i}\hat{x}_{i})$). The basis states are gaussian:
\begin{equation}
\Psi (x)\propto e^{-N \sum\limits_{i}x_{i}^{2}}
\end{equation}
Dividing in $A$ and $B$, with $A$ formed by the oscillators $\hat{x}_{i}\,;\,i=1,\cdots ,m$, and $B$ by $\hat{x}_{i}\,;\,i=m+1,\cdots ,N$\footnote{Although this partition is not gauge invariant, the $O(N)$ symmetry ensures the reduced state is the same for any partition with equal number of oscillators. Averaging over partitions provides $O(N)$ invariant subsystems with the same properties. Equivalently, the operator $x_{1}(t)x_{1}(0)$ is not gauge invariant, but its expectation value is equal to that of a gauge invariant operator $\sum\limits_{i} x_{i}(t)x_{i}(0)/N$, due to the $O(N)$ symmetry.}, we arrive at a triple of irreducible representations $\mathcal{C}=\mathcal{R}_{N}^{G}$, $\mathcal{A}=\mathcal{R}_{m}^{G} $ and $\mathcal{B}=\mathcal{R}_{N-m}^{G} $, of a unimodular coherence group $G$, satisfying $\mathcal{C}\subset \mathcal{A}\otimes\mathcal{B}$ (notice the global base state is the direct product of the local base states). For any state in the vector model $\vert\Psi\rangle \in \mathcal{C}$, and theorem~(\ref{qdefing}) asserts there exists a measure $m(g)$ on $G$ satisfying:
\begin{equation}\label{redcovec}
\Vert \textrm{Tr}_{\mathcal{B}}\vert\Psi\rangle\langle\Psi\vert -\int \vert g_{N}\rangle\langle g_{N}\vert dm (g) \Vert \leq 2 (1-\frac{d_{\mathcal{B}}}{d_{\mathcal{C}}})\;,
\end{equation}
This equals~(\ref{redco}) after identifying $m(g)$ with the diagonal entries of $e^{-H_{A}}$ in the coherent state basis.

Matrix/gauge theories are models whose degrees of freedom are matrices $M$. Dividing the system in color space is more subtle than with vector models. Reducing to submatrices of size $M$, we arrive at two irreducible representations of the coherence group, the global $\mathcal{R}^{G}_{N}$ and the local $\mathcal{R}^{G}_{M}$. The generalized theorem~(\ref{qdefingg}) applies and the reduced state is close to a convex combination of local coherent states, with the measure $m(g)$ given by the diagonal entries of $e^{-H_{A}}$ in the coherent state basis \footnote{To find the error bound we should explore all possible irreducible representations $\mathcal{R}^{G}_{M'}$ such that $\mathcal{R}^{G}_{N}\subset \mathcal{R}^{G}_{M}\otimes\mathcal{R}^{G}_{M'}$. This seems a difficult task but we notice that the error will decrease as $M$ decreases with respect to $N$.}. 

\subsection{Schmidt decomposition and entanglement entropy}

We have shown that dividing the system in $\mathcal{A}$ and $\mathcal{B}$ leads to:
\begin{eqnarray}\label{redab}
\rho_{A}=e^{-H_{A}}&\simeq & d_{A}^{2}\int\limits_{G} e^{-H_{A}}(g)    \vert g_{A}\rangle\langle g_{A}\vert   d\mu (g)\nonumber\\
\rho_{B}=e^{-H_{B}}&\simeq & d_{B}^{2}\int\limits_{G} e^{-H_{B}}(g)    \vert g_{B}\rangle\langle g_{B}\vert   d\mu (g)\;.
\end{eqnarray}
For global pure states, entanglement entropies have to equal each other for any bipartiotion:
\begin{equation}\label{sch1}
d_{A}^{2}e^{-H_{A}}(g) = d_{B}^{2}e^{-H_{B}}(g)\equiv e^{-H_{\textrm{red}}}(g) \;,
\end{equation}
and we achieve the following unifying conclusion: global states $\vert\Psi\rangle\in\mathcal{C}$ are well approximated by a `thermofield double coherent state':
\begin{equation}\label{sch2}
\vert\Psi\rangle\simeq \int\limits_{G} e^{-H_{\textrm{red}}}(g)    \vert g_{A}\rangle\otimes\vert g_{B}\rangle d\mu (g)\;.
\end{equation} 
In simple words, local coherent states provide a good approximation for the Schmidt decomposition in large-N theories. From a paralell perspective, such approximation is good due to strong decoherence in the coherent state basis.

Equations~(\ref{sch1}) and~(\ref{sch2}) imply non-trivial relations between high and low point correlations functions, analogous to what happened in SYK. Analyzing a small subsystem with high accuracy allows computing its modular Hamiltonian. But knowing the modular Hamiltonian of one subsystem, together with~(\ref{sch1}) or~(\ref{sch2}), allows the construction of the complementary reduced state, and therefore the computation of higher point correlation functions.

Notice that for spatial bipartitions in large-N theories, equation~(\ref{sch2}) is not valid. Although reduced subsystems are still convex combinations of local coherent states, each subsystem has its own coherence group.

Similar reasoning can be used in the context of eternal black holes in AdS \cite{eternal}, telling us that the thermofield double state:
\begin{equation}
\vert \textrm{TFD}\rangle \equiv \sum\limits_{n}e^{-\frac{\beta E_{n}}{2}}\vert n_{1}\rangle\otimes\vert n_{2}\rangle\simeq \int\limits_{G} e^{-H_{\textrm{red}}}(g)    \vert g_{1}\rangle\otimes\vert g_{2}\rangle d\mu (g)\;,
\end{equation}
is well approximated by local coherent states.

Finally, joining results and defining:
\begin{eqnarray}\label{op}
\mathcal{O}_{S_{A}}&\equiv & -\int\limits_{G}(\log{e^{-H_{\textrm{red}}}(g)})\vert g_{A}\rangle\langle g_{A}\vert d\mu (g)\simeq \int\limits_{G}H_{\textrm{red}}(g)\vert g_{A}\rangle\langle g_{A}\vert d\mu (g)\nonumber\\
\mathcal{O}_{S_{B}}&\equiv &-\int\limits_{G}(\log{e^{-H_{\textrm{red}}}(g)})\vert g_{B}\rangle\langle g_{B}\vert d\mu (g)\simeq \int\limits_{G}H_{\textrm{red}}(g)\vert g_{B}\rangle\langle g_{B}\vert d\mu (g)\;,
\end{eqnarray}
it follows that the expectation value of such operators provides the entanglement entropy $S_{A}=\textrm{Tr}(\rho_{A}\mathcal{O}_{S_{A}})=\textrm{Tr}(\rho_{B}\mathcal{O}_{S_{B}})=S_{B}$. More interestingly, large-N theories support big subspaces around each classical state, behaving as little perturbations around it. For such perturbations:
\begin{eqnarray}
\tilde{\rho}_{A}&\simeq &\int\limits_{G}(p_{g}+\delta_{g})\vert g_{A}\rangle\langle g_{A}\vert d\mu (g)\nonumber\\
\tilde{\rho}_{B}&\simeq &\int\limits_{G}(p_{g}+\delta_{g})\vert g_{B}\rangle\langle g_{B}\vert d\mu (g)\;,
\end{eqnarray}
so to leading order:
\begin{equation}
\tilde{S}_{A}=\textrm{Tr}(\tilde{\rho}_{A}\mathcal{O}_{S_{A}})=\textrm{Tr}(\tilde{\rho}_{B}\mathcal{O}_{S_{B}})=\tilde{S}_{B}\;.
\end{equation}
We conclude that~(\ref{op}) computes entanglement in all states close to the classical solution, and also that the deviation is linear in the perturbation from the classical state.

\section{Conclusions}

The main conclusions can be stated as follows. In theories with a classical limit, coherent states provide the preferred basis. This basis changes as the system splits. Generically, global coherent states are highly entangled mixtures of local coherent states, since the symmetry of the theory forces the global state to be part of a single irreducible representation of the coherence group. Reduced states then show strong decoherence in the coherent state basis. They are well approximated by their diagonal~(\ref{redab}), and entanglement is just the Shannon entropy of the distribution of coherent states. All aspects can be unified by stating that local coherent states universally approximate the Schmidt decomposition of bipartitions~(\ref{sch2}), implying a relation between subsystems modular Hamiltonians~(\ref{sch1}), and therefore non-trivial relations between high and low point correlation functions. This has been shown using de Finetti theorems~(\ref{cdefin}),~(\ref{qdefing}),~(\ref{qdefingg}) and independently by an argument based on the properties of coherent states in the classical limit~(\ref{global}) and~(\ref{redco}).

This approach agrees with AdS/CFT \cite{adscft} expectations. If the reduced subsystem is a sphere, the state is conformal to a thermal state in the hyperboloid \cite{marinahoracio}, and the modular Hamiltonian is a classical operator. For more generic surfaces the modular Hamiltonian is going to be a non-local operator (in CFT-space), but holographic duality predicts it to be a classical operator still, and it would be approximately diagonalized by local coherent states.

We end up with three remarks. First, our results contribute to approaches to entanglement wedge reconstruction based on modular flows \cite{aitornuevo,hayden,gabor}. Such approaches suggest that entanglement wedge reconstruction is naturally done in the eigenbasis of the modular Hamiltonian, shown here to be universally given by coherent states. Second, all these approximations are not only conceptually appealling but computationally useful, since the analytic expressions for the coherent states are known, and reduced density matrices can be explicitly constructed in the large-N limit. Finally, it would be interesting to continue the study of thermalization in large-N theories by considering Fokker-Planck type equations for the measure $m(g,t)$ of the coherence group, an approach initiated in Ref. \cite{usperm} in the SYK context. The present results agree upon and further motivate such approach to thermalization for general large-N theories, since they transparently show the strength of decoherence in the coherent state basis.

\section*{Acknowledgements}

The author is indebted to Horacio Casini and Eduardo Test\'e for comments on the draft and valuable discussions. The author also wish to acknowledge the hospitality of the Stanford Institute of Theoretical Physics, in which part of this work was developed. This work was supported by the Simons foundation through the It From Qubit Simons collaboration.




\end{document}